\documentclass[review]{elsarticle}

\usepackage{hyperref}

\journal{Acta Materialia}

\begin{document}

\begin{frontmatter}

\title{Kinetics of dislocation annealing and the effect of trapped hydrogen, investigated with \emph{in-situ} diffraction}






\author[griffith]{T.A. Webb\corref{mycorrespondingauthor}}
\cortext[mycorrespondingauthor]{Corresponding author}
\author[griffith]{C.J. Webb}
\author[griffith,ESPOL]{C.V. Tapia-Bastidas}
\author[griffith]{E.MacA. Gray}
\address[griffith]{Queensland Micro- and Nanotechnology Centre, Griffith University, Brisbane 4111, Australia.}
\address[ESPOL]{Faculty of Mechanical Engineering and Production Science, Escuela Superior Politécnica del Litoral (ESPOL), Guayaquil, Ecuador.}



\begin{abstract}
\emph{In-situ} powder diffraction was used to study the annealing of dislocations in the archetypal hydrogen absorbers Pd and LaNi$_5$. The relationship between dislocations and trapped hydrogen was explored using thermally induced desorption. It was found that the dislocations in Pd caused by hydrogen absorption anneal over a wide range of temperatures and that although they start to anneal below 250~$^\circ$C, temperatures well above 750~$^\circ$C are required to fully anneal the metal. It was shown that allowing further time at lower temperatures does not further anneal the metal. It is suggested that this is due to dislocation tangling and pinning, causing different temperatures to be required for different pinning defects. It was found that hydrogen trapped in LaNi$_5$ is released in a wide range of temperatures and it was therefore concluded that hydrogen is trapped in the dislocation strain field and dislocation core as well as vacancies. The direct comparison of deuterium release and dislocation density showed no correlation, in agreement with previous indirect comparisons. Dislocations in LaNi$_5$ were shown to anneal at temperatures as low as 150~$^\circ$C, in contrast to previous reports which suggested more than 500~$^\circ$C was required. This lower annealing temperature for dislocations at least partly explains why low temperature ageing increases the pressure hysteresis in hydrogen cycled LaNi$_5$.
\end{abstract}


\end{frontmatter}


\section{Introduction}
Each of the classic hydrogen absorbers studied in this work exhibits a very high dislocation density after absorption and desorption of hydrogen \cite{Jamieson197685,Kisi2002202,Cerny2000997}. This is a common feature of many interstitial metal hydrides and results from the misfit between the $\alpha$ and $\beta$ hydride phases. The dislocations are created at the phase boundary as it propagates through the material \cite{Jamieson197685}. The high dislocation density has been observed by Transmission Electron Microscope (TEM) imaging \cite{Jamieson197685}, neutron diffraction \cite{Kisi2002202,Wu1998363} and X-ray diffraction \cite{Wu2002229,Cerny2000997,Cerny2002288}. In some cases the dislocation density has been reported to be higher than in cold worked metals \cite{Sakaki20022652,Kisi2002202}. Pd and LaNi$_5$ have very different mechanical properties: Pd is ductile, while LaNi$_5$ is rather brittle. It is therefore expected that, relative to Pd, the dislocation width in LaNi$_5$ will be relatively large and correspond to a large volume of strained lattice .

Dislocations can be removed by several processes. They may migrate to the edge of the grain or several dislocations may align to form a new grain boundary. Externally applied stress can also affect the the movement of dislocations. If the process of dislocation annealing is activation driven and there is a given activation energy for a particular dislocation type and annealing process, then there should be corresponding characteristic temperatures at which the dislocations anneal. However, if the dislocations are pinned then the temperature of annealing will depend more on the pinning centres than the dislocation. Given that there are many different types of pinning centres \cite{Kovacs1973}, there may be a spectrum of activation energies, and therefore temperatures, at which annealing takes place.

The annealing of dislocations in hydrogen cycled Pd was investigated by \citet{Sakaki2006204} using positron annihilation spectroscopy. They found a significant drop in positron lifetime between 100~$^\circ$C and 300~$^\circ$C, which they attributed to annealing of point defects, and a further decrease in positron lifetime between 400~$^\circ$C and 650~$^\circ$C, which was attributed to annealing of dislocations. It was also suggested that the vacancies migrate to form dislocation loops and microvoids which only anneal at higher temperatures.

Hydrogen is known to be trapped in many metals and this can lead to effects such as hydrogen embrittlement. Hydrogen trapping accompanies the dislocation generation in LaNi$_5$, with about 0.08 H per metal atom (H/M) remaining after desorption into vacuum at room temperature. A much smaller quantity of hydrogen is trapped in hydrogen cycled Pd. Because of its greater dislocation width, a dislocation in LaNi$_5$ should affect a large volume of strained lattice containing many trap sites relative to Pd, which is consistent with hydrogen cycled Pd having a much smaller quantity of trapped hydrogen. There are several possible sites for hydrogen trapping in metals, mostly involving interactions with defects. \citet{Myers1992559} reviewed the interaction of hydrogen with defects. For the example of Pd, hydrogen may be trapped in vacancies, dislocation cores and in the dislocation strain field, as well as at other sites such as surface traps. The hydrogen trapping energy for a mono-vacancy is approximately 0.23 eV, approximately 0.6 eV for a dislocation core, and varies from very small values to several tenths of an eV in the strain field surrounding a dislocation \cite{Myers1992559}. Given the interaction of hydrogen with defects, the annealing properties of metals that have been hydrogen cycled and the resulting trapped hydrogen must be at least partially linked. Relating the annealing of dislocations (or any particular defect) to the release of trapped hydrogen is a difficult problem in general because of the significant variation in the trapping energies of the possible trapping sites. This variation in trapping energies can cause a wide range of temperatures at which hydrogen will desorb, and so correlating the temperature of hydrogen desorption with the temperature of dislocation annealing may not be feasible.

\citet{Sakaki20021494} studied hydrogen-cycled LaNi$_5$ using position annihilation spectroscopy and compared the annealing of vacancies to the release of hydrogen in Temperature Programmed Desorption (TPD) measurements. Their results suggested that the annealing of dislocations happened at a slightly higher temperature than the release of trapped hydrogen but that the temperature of hydrogen release roughly corresponded to the annealing of point defects (such as vacancies).

\citet{Kisi2002202} also related neutron diffraction results to \emph{ex-situ} TPD measurements of hydrogen release and noted that the main hydrogen desorption peak occurred at 500 K (227~$^\circ$C). However, their published results actually show that at 227~$^\circ$C the dislocation density had decreased by approximately 30\% and by approximately 83\% at 527~$^\circ$C. A problem with their comparison is that the \emph{ex-situ} TPD measurements were taken at a much higher ramp rate than the diffraction measurement, shifting all spectral features to higher temperature and making the comparison indirect. \citet{Kisi2002202} carried out an \emph{in-situ} neutron diffraction experiment to measure the change in dislocation density (annealing) with temperature.  It was reported that the most significant change in dislocation density happens at 800 K (527~$^\circ$C).

Vacancies are known to have a small effect on peak positions in a diffraction pattern, but do not typically have an effect on the peak breadth (nor do other point defects). However, defects can in some cases appear to behave as a different class of defect \cite{krivoglaz2011x}. For example defects of the second class can behave as first class defects if there is a shielding effect and defects of the first class can behave as though they are second class if they are present in an unusually high density. The density of vacancies in hydrogen-cycled LaNi$_5$ is reported to be exceptionally high \cite{Sakaki20021494}, well above that in thermal equilibrium. However, the effects of vacancies and dislocations are difficult to separate since there is an overlap in the temperatures at which they anneal and because they typically occur under similar circumstances.

This investigation aimed to analyse the dislocation annealing kinetics in Pd using \emph{in-situ} X-ray diffraction and authoritatively correlate the annealing of dislocations and the release of trapped hydrogen in LaNi$_5$ by simultaneously carrying out TPD and \emph{in-situ} neutron diffraction.

\section{Materials and Methods}

\subsection{Dislocation broadening}
The early work of \citet{Krivoglaz1969} provided a theoretical basis for a physical dislocation line broadening model. This was applied to hexagonal crystal systems by \citet{Klimanek198859} and put into a form appropriate for Rietveld refinement by \citet{Wu1998356}. Peak broadening due to microstrain is modelled using a Voigtian peak shape with the FWHM  given by $H=S\tan\theta$, where $S$ is the microstrain parameter. In the model published by \citet{Wu1998356} the parameter $S$ becomes a function of the dislocation density $\rho$, the orientation factor $\chi$, and the dislocation distribution parameter $M$ which is related to the range of the strain field. In the case of cubic crystals there is also the parameter $\eta$ which is related to the nature of the dislocations. The orientation factor $\chi$ is a function of the Miller indices and is different for each dislocation slip system and crystal symmetry. Formulae for the orientation factor were published by \citet{Klimanek198859}, which are referred to by subsequent publications \cite{Decamps2005570,Kisi2002202,Cerny2002288,Wu1998363,Klimanek198859}. However, there are significant discrepancies between plots of the orientation factor in the literature and so the calculations in this report use the formulae directly from \citet{Klimanek198859}. This resulted in dislocation densities which were significantly lower than those reported in the literature.

The density of dislocations in hydrogen cycled LaNi$_5$ was reported by \citet{Wu1998363} to be 4.8 $\times$ 10$^{12}$ cm$^{-2}$, based on an incorrect burgers vector, resulting in a reported density that was too high by a factor of nine. Accounting for this difference, the dislocation density would be 5.3 $\times$ 10$^{11}$ cm$^{-2}$. Other reported values (all determined using \emph{in-situ} neutron diffraction) are 3.8 $\times$ 10$^{11}$ cm$^{-2}$ (\citet{Cerny2002288}) and 5.8 $\times$ 10$^{11}$ cm$^{-2}$ (\citet{Kisi2002202}). However the current authors were not able to reproduce the plots of the orientation factor used in these reports. The dislocation density for deuterium cycled LaNi$_5$ calculated in this work is $\approx$50 times lower.

\subsection{Palladium}
The Pd sample was Goodfellow Pd powder with 99.95\% purity and 150 $\mu$m maximum particle size. The sample used had been previously cycled with deuterium. A high dislocation density was ensured by first hydrogenating at room temperature using 22 bar applied hydrogen gas pressure in a custom Sieverts apparatus, leaving under pressure for 2 hours, then desorbing by evacuating. It was then maintained under dynamic vacuum for approximately 14 hours to ensure complete desorption of diffusible hydrogen, leaving only trapped hydrogen.

The sample was then loaded onto a flat plate sample holder and measured on a PANalytical Empyrean XRD (Ag K$_\alpha$ source, divergence slits set to 1/8$^\circ$) with an Anton Paar non-ambient furnace. The furnace was evacuated and aligned prior to the measurements. An initial diffraction pattern was collected from the as-prepared sample at room temperature for 2 hours. The sample was then heated to 250~$^\circ$C for 24 hours while repeatedly collecting diffraction patterns for 2 hours each. This procedure was repeated at 500 and 750~$^\circ$C. After cooling to room temperature the sample was not fully annealed (as indicated by the diffraction pattern) and so the annealing procedure was subsequently repeated at 1000~$^\circ$C. Rietveld refinements were carried out using Topas Academic \cite{ta}. The anisotropic dislocation model mentioned above was used to fit the profiles, using a Gaussian function in preference to a full Voigtian because it reduced the number of parameters while providing a sufficient fit to the data. There was no requirement for the additional information provided by a more sophisticated profile function. The parameter $\eta$ was constrained to 1.88 as reported previously \cite{Wu1998363}. The refined dislocation parameter is proportional to the dislocation density \cite{Wu1998356,Wu1998363}.

An \emph{ex-situ} Thermal Desorption Spectroscopy (TDS) experiment was carried out on Pd for comparison with the above annealing experiment. TDS measures species-resolved desorption (with a mass spectrometer), while TPD (used below) indicates desorption measured as total pressure (with a vacuum sensor). A piece of Pd foil 5 mm square was used. The same preparation was used as for the XRD experiment. A high dislocation density was ensured by using the same preparation procedure described above for the diffraction measurement. The sample was loaded into a custom TDS apparatus, which was specifically designed to accurately measure the solubility, diffusivity and trapping of hydrogen in materials at ppm levels \cite{Bastidas_TDS}. It was then maintained under dynamic vacuum for approximately 14 hours to ensure complete desorption of diffusible hydrogen. The sample was heated at a ramp rate of 2~$^\circ$C/minute, while measuring the partial pressures of the gasses released with the mass spectrometer.

\subsection{LaNi$_5$}
LaNi$_5$ powder from Santoku Corporation was hydrogenated and then desorbed using a custom Sieverts apparatus. TPD measurements were then performed with a ramp rate of 1~$^\circ$C/min. The sample was only evacuated for a very short time (\textless 3 min) after desorption before starting the temperature ramp. The rate of hydrogen release was measured by an MKS 925 MicroPirani vacuum pressure gauge, calibrated to H$_2$ flow rate.

For the \emph{in-situ} neutron diffraction experiment, diffraction patterns were collected on HRPT (High Resolution Power diffractometer for Thermal neutrons) at the Paul Scherrer Institut, Switzerland \cite{Fischer2000146_HRPT} at a neutron wavelength of 1.494 \AA. Santoku LaNi$_5$ powder was loaded into a 7 mm ID stainless steel sample cell and fully deuterated by applying deuterium gas pressure using a custom Sieverts gas handling apparatus and evacuated to desorb. The sample was then heated at 150~$^\circ$C for approximately 5 hours, with a diffraction pattern collected during the final hour. This was repeated in 50~$^\circ$C temperature steps up to 400~$^\circ$C. The deuterium release was simultaneously measured using the equipment and method described above. The diffraction data were analysed with Topas Academic \cite{ta} using the anisotropic dislocation broadening model mentioned above. As the sample was never fully annealed (the cell was not rated to high enough temperature), a reference pattern for a fully annealed material was not possible. Some parameters were therefore constrained to obtain a stable and realistic refinement. The percentage of dislocations with basal slip planes was constrained to 10\% and the parameter $M$ constrained to 1.8. These values are based on previous literature reports \cite{Kisi2002202,Cerny2002288}.
\section{Results}
\subsection{Palladium}

\begin{figure}
\includegraphics[width=\columnwidth]{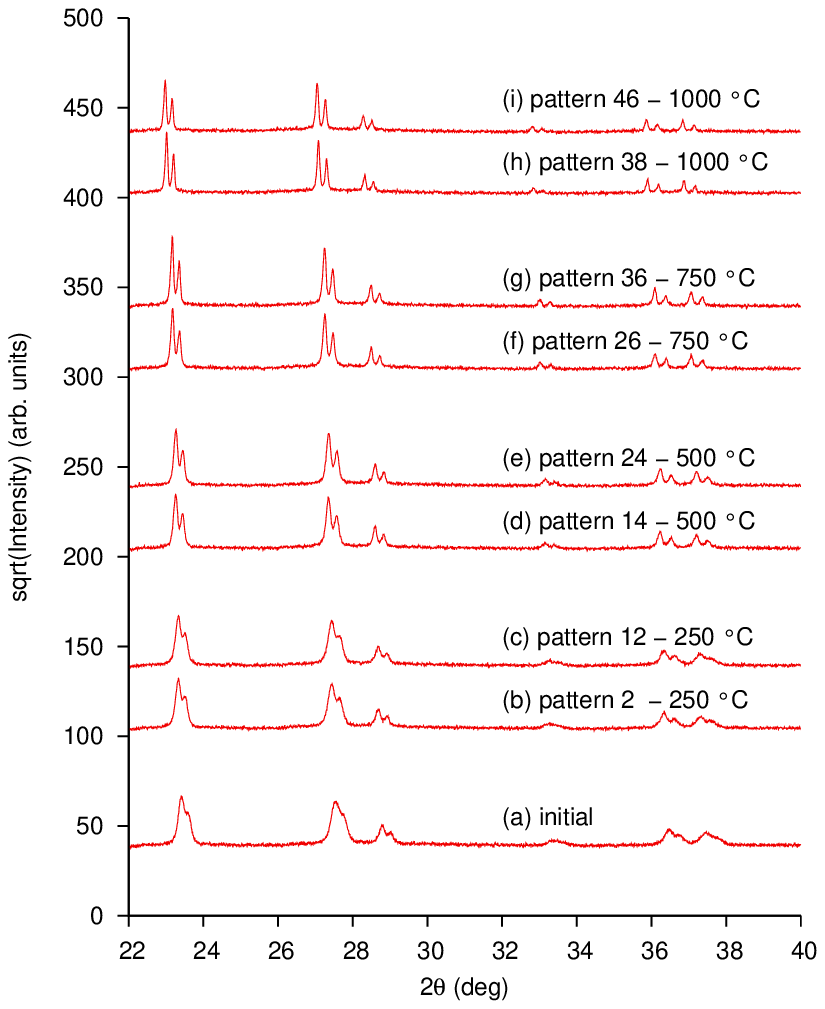}
\caption{X-ray diffraction patterns for the initial hydrogen-cycled Pd sample and during subsequent annealing. The sequence letters indicated correspond to the second and last pattern collected at each temperature.
\label{patterns}%
}
\end{figure}

\begin{figure}
\includegraphics[width=\columnwidth]{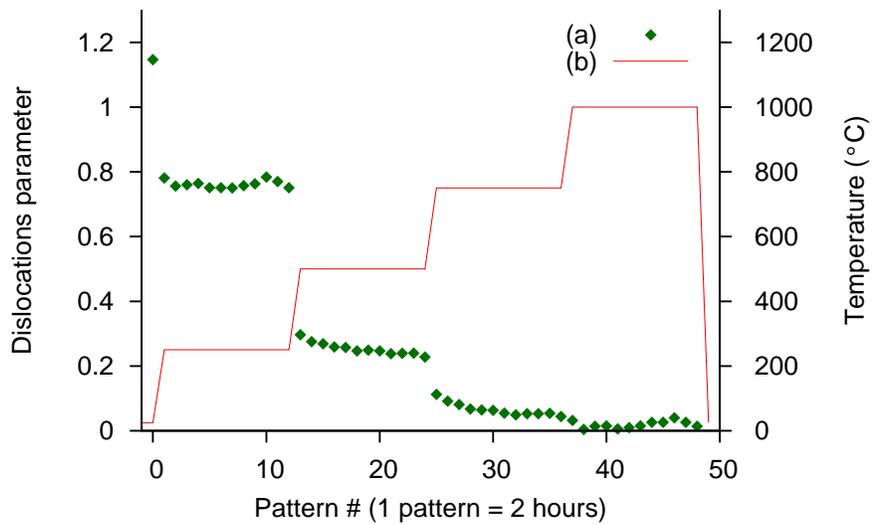}
\caption{Dislocation density parameter for Pd (a) obtained from X-ray diffraction as a function of time and temperature (b). \label{xrd}%
}
\end{figure}

\begin{figure}
\includegraphics[width=\columnwidth]{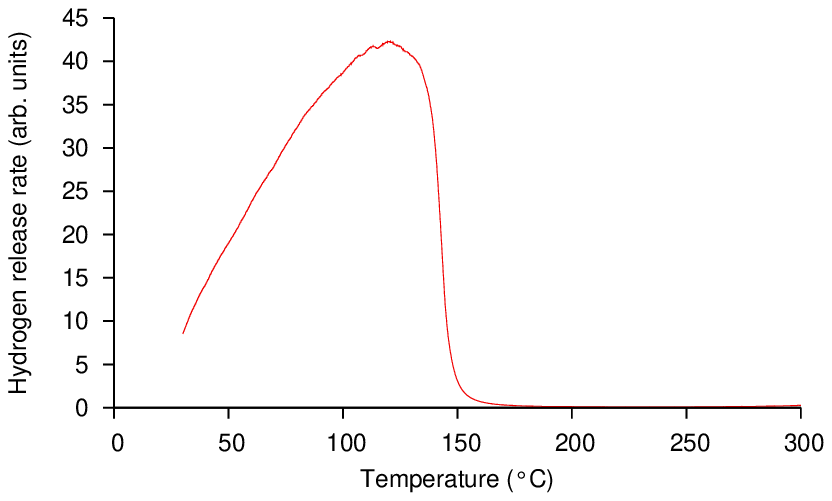}
\caption{TDS measurements on Pd carried out \emph{ex-situ} at a heating rate of 2~$^\circ$C/min. \label{pdtds}%
}
\end{figure}

X-ray diffraction patterns from the \emph{in-situ} annealing of Pd are shown in Fig. \ref{patterns}. The second pattern and the last pattern at each temperature are shown to compare the changes over time at constant temperature. The second pattern at each temperature was used rather than the first to ensure the sample had reached thermal equilibrium. The fitted dislocation parameter, which is proportional to the dislocation density, is shown in Fig. \ref{xrd}. This shows that even at the lowest temperature of 250~$^\circ$C the sample reached a significantly lower dislocation density compared to room temperature. This agrees with the authors' earlier work\cite{Webb2015_PRB}. Interestingly the dislocation density did not continue to decrease over 24 hours at this temperature; a steady state was reached during the first 2-hour period. At 500~$^\circ$C the dislocation density decreased to approximately a quarter of that of the initial cycled sample. At this temperature there was some continued annealing with time, but at an extremely low rate (approximately 10\% over 24 hours). At 750~$^\circ$C the dislocation density again dropped, this time with obvious continued annealing with time, although still at a very low rate. The data at 1000~$^\circ$C were difficult to fit because the sample sintered, which significantly increased its packing density and resulted in the sample no longer being flush with the top of the sample holder. This change in sample position and shielding of the sample from the beam  by the sample holder wall at low angles had a significant effect on the peak positions, intensities and peak shape. The positions and intensity were corrected as well as possible in the Rietveld analysis, but it was not possible to adequately correct for the peak shape. Therefore the dislocation density parameter at this temperature is less reliable. The sintering of the sample also made it impossible to obtain a high quality diffraction pattern at room temperature after annealing.

The \emph{ex-situ} TDS measurements of the trapped hydrogen in Pd are shown in Fig. \ref{pdtds}. Essentially all the hydrogen was released below 150~$^\circ$C, whereas the greatest decrease in dislocation density occurred after heating to 500~$^\circ$C. This establishes that for hydrogen cycled Pd there is no direct correlation between hydrogen release and dislocation annealing.

\subsection{LaNi$_5$}

\begin{figure}
\includegraphics[width=\columnwidth]{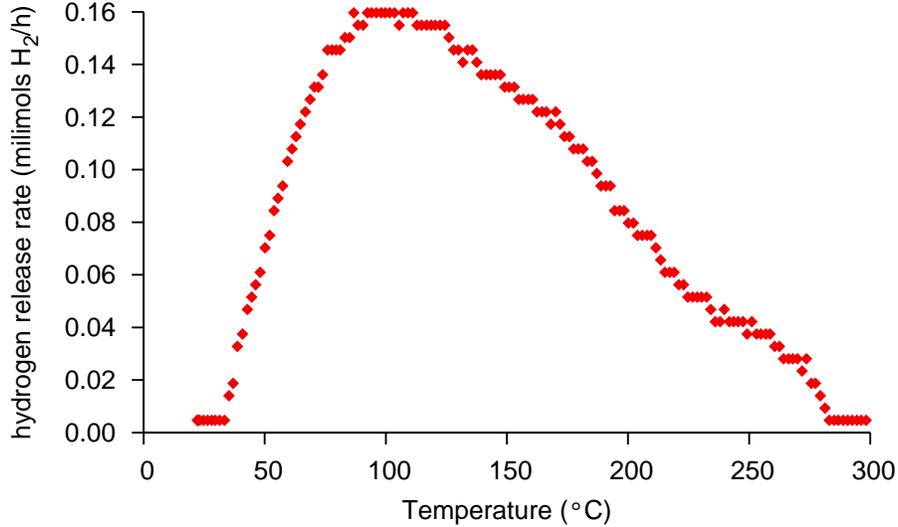}
\caption{{TPD measurements on LaNi$_5$ carried out \emph{ex-situ} at a heating rate of 1~$^\circ$C.
\label{labtpd}%
}}
\end{figure}

\begin{figure}
\includegraphics[width=\columnwidth]{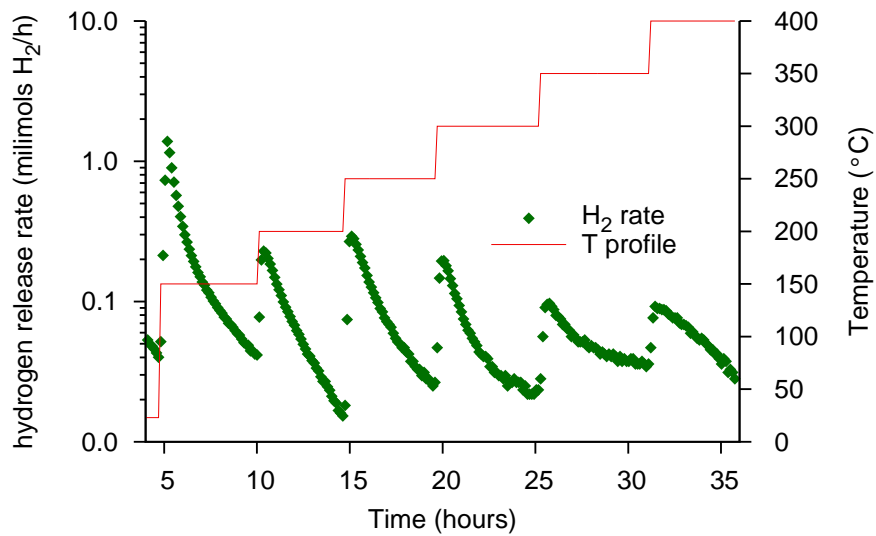}
\caption{{LaNi$_5$ TPD measurements made simultaneously to \emph{in-situ} neutron diffraction.
\label{insitutpd}%
}}
\end{figure}

\begin{figure}
\includegraphics[width=\columnwidth]{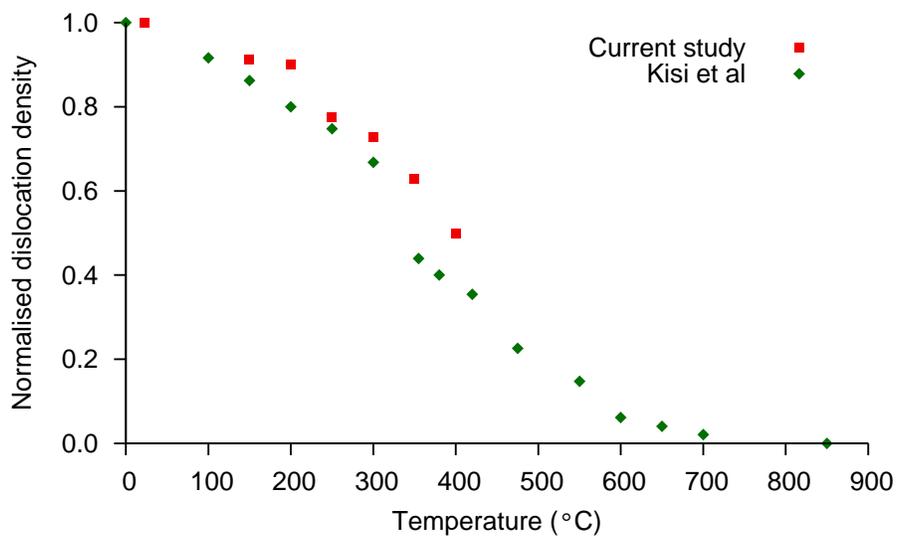}
\caption{{Dislocation density in hydrogen-cycled LaNi$_5$ measured using neutron diffraction, compared with the results of \citet{Kisi2002202}, normalised to the initial value at room temperature.
\label{lani5annealing}%
}}
\end{figure}

Figure \ref{labtpd} shows the results of the \emph{ex-situ} TPD measurement for LaNi$_5$. Hydrogen desorption both starts and finishes at lower temperatures than  reported previously \cite{Sakaki20021494,Kisi2002202}, which is expected given the much slower temperature ramp used in the current study (1~$^\circ$C/min used here, while \citep{Sakaki20021494} used 12~$^\circ$C/min for TPD and 120~$^\circ$C/min for TDS). However, the hydrogen release started at about 40~$^\circ$C, whereas previous reports suggested that hydrogen was not released until approximately 150~$^\circ$C \cite{Sakaki20021494,Kisi2002202}, although TDS measurements suggested that small amounts of hydrogen were released just above room temperature \cite{Sakaki20021494}. This can partly be explained by the difference in temperature ramp rates and partly by the initiation of the temperature ramp within 3 min of placing the sample under vacuum in our study. Therefore if some of the hydrogen would have been released at room temperature, but with very slow kinetics, it would not have had the chance to do so before the start of the temperature ramp. This may have added to the hydrogen release at relatively low temperature. The TPD profile shows a single, very broad peak for hydrogen desorption, rather than the distinct but heavily overlapped two peaks observed previously \cite{Sakaki20021494,Kisi2002202}.

The stepped TPD spectra obtained during the \emph{in-situ} annealing experiment are shown in Fig. \ref{insitutpd} and the corresponding dislocation densities are shown in Fig. \ref{lani5annealing}, normalised for comparison. The dislocation density of cycled LaNi$_5$ was calculated to be 8.8 $\times$ 10$^{10}$ cm$^{-2}$. The trend of the dislocation density during annealing was very similar to that observed by \citet{Kisi2002202}, although the numerical values are different for the reasons given in the Materials and Methods.

The \emph{in-situ} TPD measurements shown in Fig. \ref{insitutpd} indicate that after heating to 150~$^\circ$C a large amount of deuterium was released from the sample, but that after subsequent temperature steps significant further amounts were released. Noting the logarithmic vertical axis, a straight line indicates an exponential decay with time of the deuterium release rate, which in turn implies a constant probability per unit time of detrapping.

There is no obvious correlation between the dislocation density (Fig. \ref{lani5annealing}) and the release of deuterium (Fig. \ref{insitutpd}). This agrees with the conclusion of \citet{Kisi2002202}, despite the different temperature ramp rate used.

An attempt was made to observe any broadening in the \emph{in-situ} neutron diffraction patterns of LaNi$_5$ which may have been associated with vacancies. However, given that dislocations anneal at a lower temperature than previously thought (probably overlapping with the vacancy annealing temperature) and that the dislocations with basal slip system cause almost isotropic peak broadening, it was not possible to distinguish any broadening which could not be explained by dislocations. Peak broadening due to vacancies in hydrogen cycled metals has not been reported in the literature.

\section{Discussion}

The annealing results for hydrogen cycled Pd show that (i) the dislocations do not anneal at a single temperature and (ii) at any given temperature some of the dislocations will anneal relatively quickly but no further dislocation annealing will happen until the temperature is increased. The dislocations in fact anneal over a wide range of temperatures, indicating that there is a range of activation energies. Hence there is no characteristic temperature for the annealing of dislocations in hydrogen cycled Pd. This is unsurprising, since the annealing of dislocations is often dependent on the defects that are pinning the dislocations and the interactions of the defects with the dislocations, rather than the energies of the dislocations themselves \cite{Kovacs1973}. This contradicts the conclusion of \citet{Sakaki2006204} that dislocations anneal above approximately 400~$^\circ$C and that at lower temperatures it is vacancies and vacancy clusters that anneal. Dislocations in fact anneal from a very low temperature, although at these low temperatures it could be that dislocations are being pinned by vacancies or other point defects and so become mobile at the same temperatures as these defects. It is also possible that there is a high vacancy concentration which anneals at lower temperatures and may not be observed in the diffraction data. The TDS measurements showed that all the trapped hydrogen is released below 200~$^\circ$C. There is therefore no direct correlation between the dislocation annealing and the hydrogen release, implying that it cannot be trapped hydrogen pinning the dislocations.

The previously reported temperature for the initiation of dislocation annealing in LaNi$_5$ based on positron lifetime studies (800 K , 523~$^\circ$C) \cite{Sakaki20021494,Shirai2002125} appears to be much too high. Although \citet{Kisi2002202} reported that dislocations anneal at 800 K (523~$^\circ$C) based on neutron diffraction, their results for the dislocation density show the same trend as our results (Fig. \ref{lani5annealing}), with a decrease in dislocation density of about 30\% by 227~$^\circ$C and 83\% by 527~$^\circ$C. However, temperatures well over 500~$^\circ$C are nevertheless required for complete annealing of LaNi$_5$.

It has been reported that very low ageing temperatures affect the plateau pressure and pressure hysteresis in LaNi$_5$ \cite{Buckley1995460}. This was unexpected, since it was thought that dislocations did not anneal below approximately 500~$^\circ$C and it was therefore suggested that the annealing of vacancies may affect the pressure hysteresis \cite{Buckley1995460}. However, given our demonstrations that dislocations start to anneal at much lower temperatures, down to at least 150~$^\circ$C, this result is much less surprising. It was, however, reported that even extremely low ageing temperatures (45~$^\circ$C) had a small but measurable effect on the hysteresis \cite{Buckley1995460}. Given that only a very small reduction in dislocation density was observed at 150~$^\circ$C, it is unlikely that a difference in hysteresis after heating to 45~$^\circ$C would be due to dislocations, so this aspect remains to be understood.

\citet{Sakaki20021494} calculated that the vacancy concentration in hydrogen cycled LaNi$_5$ was between 10$^{-3}$ and 10$^{-1}$, well above the equilibrium vacancy concentration in metals at the melting point which is between 10$^{-5}$ and 10$^{-4}$. However, the calculations were based on the incorrect density of dislocations in hydrogen cycled LaNi$_5$ of 4.8 $\times$ 10$^{12}$ cm$^{-2}$ calculated by \citet{Wu1998363}. The dislocation density for deuterium cycled LaNi$_5$ found in the present work is $\approx$50 times lower. This means that the dislocation density used by \citet{Sakaki20021494} to calculate the vacancy concentration was too high by 1-2 orders of magnitude and therefore the calculated vacancy concentration was correspondingly too high.

It has been proposed that the trapped hydrogen/deuterium sits in vacancies, partially stabilising the vacancies and therefore allowing for the unusually high vacancy concentration in hydrogen cycled metals \cite{Myers1992559}. Given that there is no characteristic temperature for hydrogen release and therefore no single activation energy, there must be range of trapping sites for hydrogen. Therefore the hydrogen cannot be trapped only in mono-vacancies. The roughly exponential decrease with time of the deuterium release rates in Fig. \ref{insitutpd} indicate that a narrow range of trap energies is explored at each temperature in the stepped TPD sequence, with the probability per unit time of escaping the trap ideally proportional to the slope of the line. The different behaviour above 300~$^\circ$C is consistent with a different type of trap. \citet{Sakaki20021494} reported that the vacancies were di-vacancies and higher order vacancy clusters and that the annealing temperature of the di-vacancies correlated with the release of trapped hydrogen. The wide range of temperatures at which deuterium was released in the \emph{in-situ} annealing (Fig. \ref{insitutpd}) suggests that although there may be deuterium in di-vacancies, there must also be other trapping sites. The quantity of trapped hydrogen in LaNi$_5$ is approximately 0.08 H/M \cite{Gray199957} and so it is also unlikely that it would be possible for all of the trapped hydrogen to lie in vacancies, since that would require a vacancy concentration on the order of 10$^{-1}$. Given the high density of dislocations and the wide variation of trapping energies for various trapping site in dislocation strain fields and cores, it is highly likely that a significant amount of hydrogen is trapped in and around the dislocations. This does not imply that the release of trapped hydrogen should occur at the same temperature as the annealing of the dislocations, although the trapped hydrogen in or around a dislocation would not persist after the annealing of the dislocation.

\section{Conclusions}
It was found that dislocations in Pd anneal over a wide range of temperatures and that, although they start to anneal at very low temperatures (\textless 250~$^\circ$C), temperatures well above 750~$^\circ$C are required to fully anneal the metal. It was shown that allowing further time at lower temperatures does not further anneal the metal. This important result provides information about the annealing process of dislocations created during hydrogen cycling. We postulate that dislocation tangling and dislocation pinning may be the reason for this wide range of temperatures in which dislocations will anneal, although this cannot be due to trapped hydrogen, since there was no direct correlation between the temperature range within which trapped hydrogen was released and the much wider range over which dislocation annealing occurred.

It was found that hydrogen trapped in LaNi$_5$ is released in a wide range of temperatures, implying a range of trapping energies and probably a range of trap types. It appears that hydrogen is trapped in sites within the dislocation strain field and dislocation core as well as in vacancies in relatively unstrained lattice. Direct comparison of released deuterium and dislocation density showed that there is no correlation between temperatures of dislocation annealing and deuterium release. This was also true of trapped hydrogen in Pd. Therefore the conclusions from previous indirect comparisons are confirmed to be correct.

Dislocations in LaNi$_5$ were shown to anneal at temperatures as low as 150~$^\circ$C. This is in contrast to previous reports which suggested that temperatures over 500~$^\circ$C were required. However, the dislocation annealing plot follows the same trend as that previously reported, despite different conclusions about the annealing temperature. This lower annealing temperature for dislocations at least partly explains why low temperature ageing increases the pressure hysteresis in hydrogen cycled LaNi$_5$.

\section*{Acknowledgments}
The authors thank D. Sheptyakov for his assistance with neutron diffraction experiments at the Swiss spallation neutron source SINQ. T.A. Webb acknowledges financial support from the Australian Institute of Nuclear Science and Engineering and receipt of an Australian Postgraduate Award (APA). The experimental work was supported by grants from the Australian Research Council.

\section*{References}

\bibliography{phd}

\end{document}